# Interdiffusion in epitaxial ultrathin $Co_2FeAl$/MgO heterostructures with interface-induced perpendicular magnetic anisotropy

Zhenchao Wen,[1, a)] Jason Paul Hadorn,[1] Jun Okabayashi,[2] Hiroaki Sukegawa,[1, b)] Tadakatsu Ohkubo,[1] Koichiro Inomata,[1] Seiji Mitani,[1,3] and Kazuhiro Hono[1,3]

[1]*Research Center for Magnetic and Spintronic Materials, National Institute for Materials Science (NIMS), 1-2-1 Sengen, Tsukuba 305-0047, Japan*
[2]*Research Center for Spectrochemistry, The University of Tokyo, Bunkyo-ku, Tokyo 113-0033, Japan*
[3]*Graduate School of Pure and Applies Sciences, University of Tsukuba, Tsukuba 305-8577, Japan*

## Abstract

The structures of epitaxial ultrathin $Co_2FeAl$/MgO(001) heterostructures relating to the interface-induced perpendicular magnetic anisotropy (PMA) were investigated using scanning transmission electron microscopy, energy dispersive x-ray spectroscopy, and x-ray magnetic circular dichroism. We found that Al atoms from the $Co_2FeAl$ layer significantly interdiffuse into MgO, forming an Al-deficient Co-Fe-Al/Mg-Al-O structure near the $Co_2FeAl$/MgO interface. This atomic replacement may play an additional role for enhancing PMA, which is consistent with the observed large perpendicular orbital magnetic moments of Fe atoms at the interface. This work suggests that control of interdiffusion at ferromanget/barrier interfaces is critical for designing an interface-induced PMA system.

[a)] Current affiliation: Institute for Materials Research, Tohoku University, Sendai, Japan

[b)] Corresponding author: sukegawa.hiroaki@nims.go.jp

Achieving large perpendicular magnetic anisotropy (PMA) in thin films using well-designed ferromagnetic materials and/or multilayer stacks is of particular importance to develop high-performance spintronic devices such as spin-transfer torque (STT) magnetoresistive random access memories (MRAMs) and magnetic non-volatile logics.[1-3] The PMA originating from the interface between a ferromagnetic (FM) metal and an oxide, due to the hybridization of orbitals between the FM and oxygen atoms, has been attracting much attention because of the prospective applications in STT-MRAMs and electric-field controlled spintronic devices.[4-15] To date, perpendicularly magnetized ultrathin films have been reported for ultrathin films of fcc-Co,[4,5] bcc-Fe,[6-8] CoFeB,[10,11] and Co-based Heusler compounds ($Co_2FeAl$,[12,13] $Co_2FeSi$,[16] $Co_2FeGe$,[17] and $Co_2Fe_{1-x}Mn_xSi$[18]) with $AlO_x$ or MgO interfaces.[19,20] Large out-of-plane tunneling magnetoresistance (TMR) ratios of more than 100% at room temperature (RT) were reported in magnetic tunnel junctions (MTJs) with CoFeB/MgO [11] and $Co_2FeAl$ (CFA)/MgO [13] structures.

In order to control PMA in an ultrathin FM/oxide structure for specific applications, clarifying the deciding factor of interfacial PMA in well-defined epitaxial structures is indispensable since PMA energy is mainly determined by the local crystal structure of a few monolayers near the FM/oxide interface[21-23] and interdiffusion at the interfaces.[24, 25] When an ordered alloy is used as an FM layer, the atomic arrangement and chemical ordering near the interface may significantly affect the PMA characteristics. Epitaxial heterostructures consisting of CFA(001) and MgO(001) layers could be good candidates for examining such interfacial structures due to the relatively small lattice mismatch between CFA and MgO,[26] and the reported large PMA at their interface.[12,13] However, the accurate atomic arrangement near the CFA/MgO(001) interfaces exhibiting large interfacial PMA is still unknown. Therefore, determination of the crystal structure and elemental distribution at these interfaces using high resolution nanostructure analyses is of particular importance to elucidate the origin of the interfacial PMA.

In this work, we investigated the interfacial atomic structure of the epitaxial CFA/MgO(001) PMA heterostructures on Cr and Ru buffer layers using high resolution scanning transmission electron microscopy (STEM) and energy dispersive x-ray spectroscopy (EDS). STEM images and EDS elemental profiles revealed Al interdiffusion at the CFA/MgO interface regardless of which Cr or Ru buffer layer was used. Notably, significant Al interdiffusion from CFA into MgO resulted in the formation of an Al-deficient CFA layer and an Mg-Al-O barrier. This interdiffusion can promote the hybridization between Fe $3d_{z^2}$ and O $2p_z$ orbitals at the interface, which is consistent with an anisotropic orbital magnetic moment of Fe evaluated by angular-dependent x-ray magnetic circular dichroism (XMCD) measurements. This study reveals that the reconstruction of the atomic arrangement by the interdiffusion plays an important role in PMA characteristics at a ferromanget/oxide interface.

Epitaxial multilayers with the structures of Cr (40)/CFA (1.2)/MgO (1.8) (see the inset of Fig. 1 (a)) and Ru (40)/CFA (1.2)/MgO (1.8) (see the inset of Fig. 1 (b)) (unit in nm) were deposited on MgO(001) single crystalline substrates by ultra-high vacuum magnetron sputtering system at the base pressure of around $3 \times 10^{-7}$ Pa. These structures were optimized to achieve both large PMA and TMR values simultaneously.[13,27] After deposition, the samples were post-annealed at $T_{ex}$ = 325°C for 30 min in a vacuum furnace. Figures 1 (a) and (b) show out-of-plane and in-plane magnetic hysteresis loops of the Cr-buffered and Ru-buffered samples, respectively, measured using a vibrating sample magnetometer at RT. The effective magnetic anisotropy energy density ($K_{eff}$) of the films was simply estimated by the equation, $K_{eff} = M_s H_k / 2$, where $M_s$ is the saturation magnetization and $H_k$ is the anisotropy field of the hard magnetization direction (therefore, a positive $K_{eff}$ indicates perpendicular magnetization). An in-plane magnetization of $K_{eff} = -1.7 \times 10^6$ erg/cm$^3$ was observed for the Cr-buffered sample, while $K_{eff} = 2.2 \times 10^6$ erg/cm$^3$ was observed for the Ru-buffered sample. This difference is mainly ascribed to the difference in the interface anisotropy $K_s$, in which the $K_s$ values were derived by assuming the following simple relationship, $K_s = (K_{eff} - K_v + 2\pi M_s^2)t$, where $2\pi M_s^2$ and $K_v$ are the shape anisotropy and bulk magnetocrystalline anisotropy energy densities, respectively, and $t$ is the thickness of the effective magnetic layer.[10] The $K_s$ values were determined to be 1.0 and 2.1 erg/cm$^2$ for the Cr- and Ru-buffered samples, respectively, which are consistent with previous studies.[12,13]

Thin foil specimens for STEM observation were prepared by the lift-out technique using an FEI Helios Nanolab 650 focused ion beam. Microstructural characterization was performed by TEM (Titan G2 80-200) with high resolution EDS element mapping capability. For the Ru-buffered sample, x-ray absorption spectroscopy (XAS) and XMCD measurements were carried out at BL-7A in the Photon Factory, High-Energy Accelerator Organization (KEK). The total electron yield mode was adopted using an applied magnetic field of ±1 T along the incident polarized soft x-ray direction during measurements at RT.

Figures 2 (a) and (c) [(b) and (d)] show the low magnification and high resolution cross-sectional annular dark field STEM (ADF-STEM) images of the Cr-buffered sample [Ru-buffered samples], respectively, along the MgO[100] || CFA[110] direction. The images indicate that the CFA and MgO layers are continuous and that they grew epitaxially on both buffers. Schematics showing the epitaxial relationships are shown in Fig. 2 (e). The Ru buffer crystallizes into an hcp structure with a nearly $(02\bar{2}3)$ interfacial orientation as in the previous report.[13] The spacing of interface-normal Ru atomic planes along the in-plane MgO{100} (∥CFA{110}) directions is approximately equal to half of the diagonal distance of the square-like structure of the Ru ($02\bar{2}3$) plane as shown in Fig. 2 (e). Figures 2 (f) and (g) show nano-beam electron diffraction (NBD) patterns taken at the CFA/MgO interfacial region of the Cr-buffered and the Ru-buffered samples, respectively. A sputter-deposited CFA film generally has a *B*2 structure. However, the *B*2 superlattice reflection spots (e.g., {002}, the Miller index for an $L2_1$ lattice) for CFA were not clearly observed for both the samples, implying a compositional change in the CFA layers and/or a reduced *B*2 order parameter, as will be discussed later.

Figures 3 (a) and (b) show EDS elemental maps of the Cr- and Ru-buffered samples, respectively, which were each taken from the corresponding region indicated by the ADF-STEM image (left of Figs. 3 (a) and (b)). Normalized line profiles of the elemental maps are shown in Figs. 3 (c) and (d); the vertical dash lines indicate the gravity points of five elements (Co, Fe, Al, Mg, and O) based upon their fitted Gaussian distribution. For each sample, the gravity position of Al appears shifted from the Co and Fe positions towards and into the MgO layer region, indicating Al diffusion into the barrier. Therefore, the absence of the *B*2 superlattice reflection spots in Figs. 2 (f) and (g) is mainly attributed to the Al depletion in the CFA layers regardless of the buffer material. The Mg peak is shifted to the right of O and farther away from the CFA interface, indicating that Al partially replaces Mg in the MgO layer and more strongly at ion sites nearer to the interface, i.e., creating a concentration gradient. The underside of the MgO layer is cation-disordered $MgAl_2O_4$ due to the absence of the spinel superlattice reflection spots.[28] Such Al interdiffusion alters the interface electronic structure significantly; thus, the PMA energy at the interface may be effectively enhanced. Additionally, we cannot exclude the possibility of a small amount of interdiffusion of Cr and Ru atoms into CFA, which may also have an unignorable impact on the PMA properties.

We also confirmed the presence of Al interdiffusion by the EDS profiles even before post-annealing as shown in Fig. 4 (a) for the Cr-buffered case and in Fig. 4 (b) for the Ru-buffered case. This fact indicates that this phenomenon proceeds during the MgO layer deposition, presumably due to the greater thermodynamic stability of CoFe/Mg-Al-O compared to CFA/MgO when a CFA layer is very thin. The Al peak was closer to the CFA/MgO interface compared to that in the post-annealed samples, thereby indicating that post-annealing further promotes Al interdiffusion.

The interfacial PMA in the Cr-buffer samples is determined mainly by the contribution of large orbital magnetic moments of Fe atoms rather than Co atoms at the CFA/MgO interface, which was experimentally confirmed by angular-dependent XMCD.[29] Here, we similarly evaluated anisotropy of orbital magnetic moments for Fe and Co for the Ru-buffered sample. Figures 5 (a) and (b) show the x-ray absorption spectra of Fe and Co $L$-edges for an annealed Ru/CFA(0.8 nm)/MgO heterostructure measured in normal incidence (NI) geometry, where the absorption processes involve the normal direction components of the orbital angular momentum ($m_{\mathrm{orbital}}^{\perp}$). Distinct metallic XAS peaks of Fe and Co $L_{2,3}$ edges are observed. The clear differences between right ($\mu^+$) and left ($\mu^-$) hand x-rays reveal the XMCD signals. Figures 5 (c) and (d) show the XMCD spectra of the NI and grazing incidence (GI) setups for the Fe and Co $L$ edges. The GI configuration mainly detects the in-plane orbital momentum components ($m_{\mathrm{orbital}}^{\parallel}$). A greater asymmetric XMCD signal for the Fe $L$-edge in the NI direction was observed compared to that in the GI direction, indicating that large orbital magnetic moments exist in the perpendicular direction to the film plane. In addition, the XMCD signal for Co $L$-edge only shows a slight deviation in both NI and GI geometries. Figures 5 (e) and (f) show the integrated XMCD spectra of the Fe and Co $L$-edges for both NI and GI geometries. The integrated value is proportional to the orbital magnetic moment indicated by the magneto-optical sum rule.[30] The residual of the integrals of the Fe *L*-edge XMCD signals between NI and GI configurations is much larger than that of the Co *L*-edge signals, indicating large perpendicular orbital magnetic moments of Fe atoms. The values of $m_{\text{orbital}}^{\perp}$ and $m_{\text{orbital}}^{\parallel}$ for Fe (Co) are estimated to be 0.31$\mu_B$ (0.19$\mu_B$) and 0.22$\mu_B$ (0.17$\mu_B$), respectively, using magneto-optical sum rules. These results suggest that the large perpendicular orbital magnetic moments of Fe mainly contribute to the interfacial PMA in the Ru-buffered case, which is similar to the Cr-buffered case.[29]

As discussed earlier, the ultrathin CFA (~ 1 nm) and MgO layers transform into a CoFe-rich layer and an Mg-Al-O barrier, respectively, regardless of the buffer material; this could induce the high interfacial PMA energies through the promotion of the hybridization between Fe $3d_{z^2}$ and O $2p_z$ orbitals. Therefore, Al redistribution at the CFA/MgO interface and its interdiffusion into the MgO layer are particular features in the systems with ultrathin CFA films. In addition, this indicates that the peculiar electronic structure of a CFA alloy, namely the half-metallic feature, cannot maintain at the interfaces in the present samples. At the same time, however, the interfacial diffusion, which is well-controlled by the deposition and post-annealing conditions, can assist to establish strong interfacial PMA using Heusler alloy based films. Meanwhile, the origin of the difference in $K_s$ values between the Cr-buffered sample (~1 erg/cm$^2$)[12] and the Ru-buffered one (~2 erg/cm$^2$)[13] is not clear at this moment. The differences in the contributions of the buffer interdiffusion and lattice distortion near the CFA layer owing to the interfacial lattice mismatch between CFA and the buffer layers may be responsible for this difference. Further systematic studies are needed to clarify the critical factor of the interfacial PMA.

In summary, the interfacial structure and interdiffusion in epitaxial ultrathin CFA/MgO heterostructures on a Cr or Ru buffer layer, relevant to interfacial PMA, were examined by ADF-STEM imaging and EDS elemental mapping. Al interdiffusion from CFA into MgO were clearly observed in CFA/MgO heterostructure, which may play an additional role for establishing PMA at the oxide interfaces. XMCD measurements revealed that the large perpendicular orbital magnetic moments of Fe at the interface mainly contribute to the PMA for both the buffered cases. This work provides a pathway for understanding and engineering a useful PMA system for future spintronic applications through consideration of interdiffusion between FM/MgO interfaces.

This work was supported by the ImPACT Program of Council for Science, Technology and Innovation, Japan. We thank Jun Uzuhashi for his technical assistance in TEM specimen preparations.

FIG. 1. Out-of-plane and in-plane magnetic hysteresis loops of epitaxial (a) Cr (40)/CFA (1.2)/MgO (1.8) and (b) Ru (40)/CFA (1.2)/MgO (1.8) (unit: nm) structures deposited on MgO(001) substrates. Both samples were annealed at $T_{ex}$ = 325°C. Insets of (a) and (b) show schematic illustrations of the respective samples.

FIG. 2. (a)-(d) Cross-sectional ADF-STEM images of Cr/CFA/MgO ((a) and (c)), and Ru/CFA/MgO structures ((b) and (d)) annealed at $T_{ex}$ = 325°C along the MgO[100] direction. (e) Schematics of the epitaxial relationships for Cr (right) and Ru (left) buffered cases. (f) and (g) NBD patterns near the CFA/MgO interface for (f) Cr/CFA/MgO and (g) Ru/CFA/MgO structures.

FIG. 3. (a) and (b) Cross-sectional STEM images and corresponding EDS maps of Cr (Ru), Co, Fe, Al, Mg, and O elements for the (a) Cr/CFA (1.2)/MgO (1.8) and (b) Ru/CFA (1.2)/MgO (1.8) (unit: nm) structures annealed at $T_{ex}$ = 325°C. (c) and (d) Normalized line averaged profiles of the elemental maps (solid line) for the (c) Cr/CFA/MgO and (d) Ru/CFA/MgO structures. The scan areas are indicated by the rectangles in the STEM images of (a) and (b). Dotted vertical lines of (c) and (d) are center positions of Gaussian fit profiles for each element profile. The numbers next to each element show the distances from the Co center positions.

FIG. 4. (a) and (b) Normalized line averaged profiles of the elemental maps (solid line) for as-deposited (a) Cr/CFA (1.2)/MgO (1.8) and (b) Ru/CFA (1.2)/MgO (1.8) (unit: nm) structures.

FIG. 5. (a) and (b) X-ray absorption spectra of (a) Fe and (b) Co $L$-edges for an epitaxial Ru/CFA (0.8 nm)/MgO (2.0 nm) structure annealed at $T_{ex}$ = 325°C measured in NI geometry. $\mu^-$ (red dotted lines) and $\mu^+$ (blue solid lines) denote the different helicities of the x-ray. (c) and (d) XMCD spectra of the NI (blue solid lines) and GI (red solid lines) setups for (c) Fe and (d) Co $L$-edges. (e) and (f) Integrated XMCD spectra for (e) Fe and (f) Co $L$-edges in both NI and GI setups. Insets in (c) and (d) are the expanded views near the $L_3$-edge XMCD.

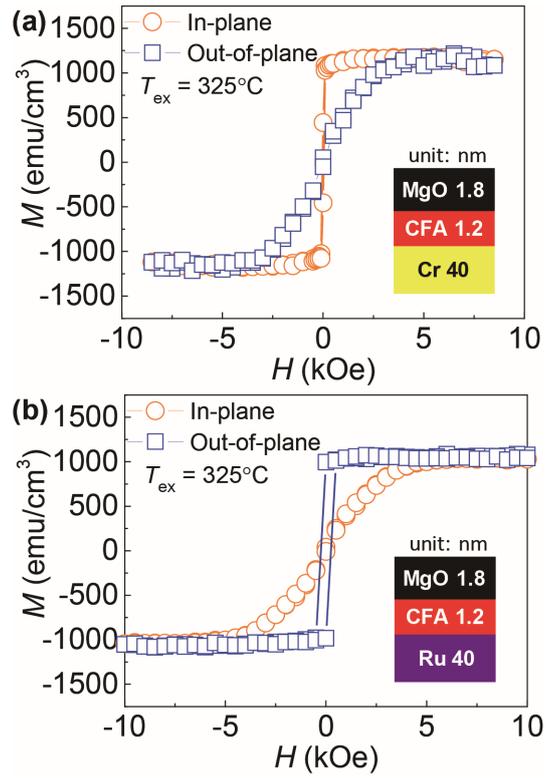

FIG. 1.

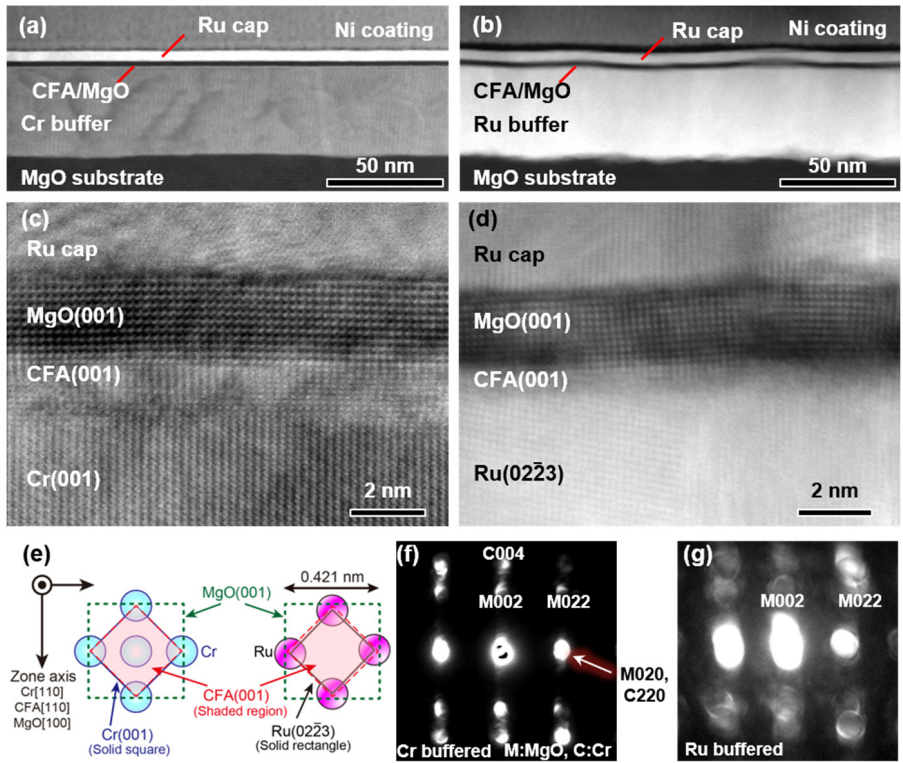

FIG. 2.

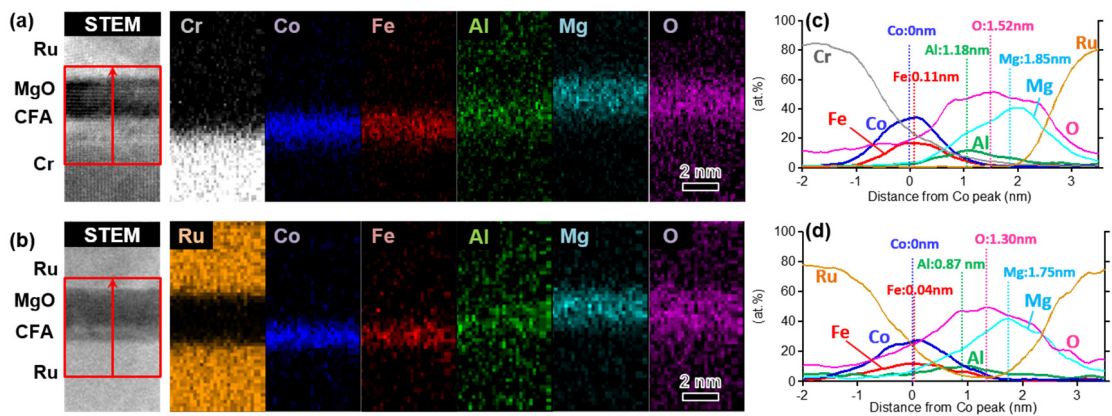

FIG. 3.

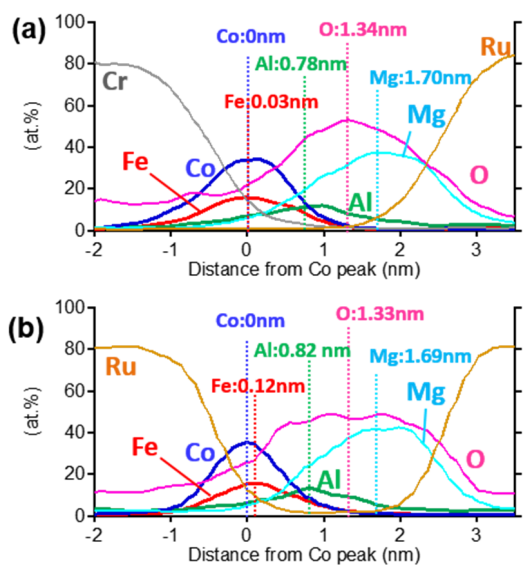

FIG. 4.

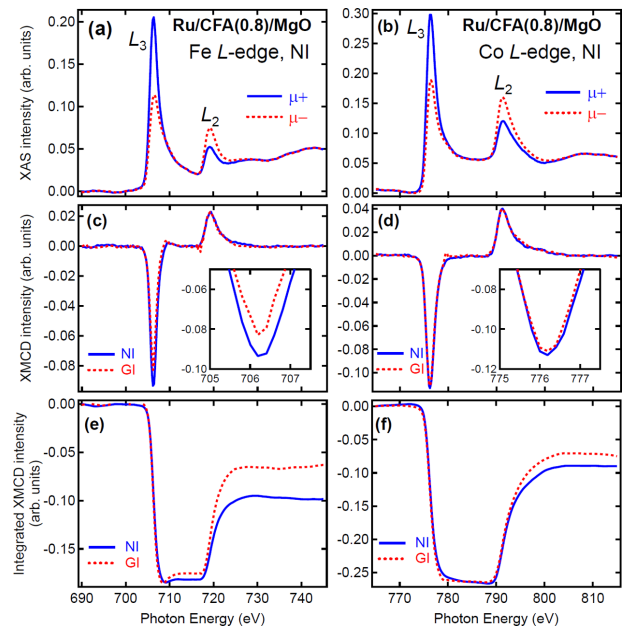

FIG. 5.